\documentclass[journal=apchd5,manuscript=letter]{achemso}

\usepackage{chemformula} 
\usepackage[T1]{fontenc} 
\usepackage{graphicx,color}
\usepackage{dcolumn}
\usepackage{bm}
\usepackage{epstopdf}
\usepackage{verbatim}
\usepackage{enumerate}
\usepackage{textcomp}

\usepackage[normalem]{ulem}   

\usepackage[official]{eurosym}
\usepackage[squaren]{SIunits}
\usepackage{xcolor}

\usepackage{color}
\usepackage{amsmath}
\usepackage{achemso}

\newcommand{\sss}{\sigma_{\mathrm{s}}}
\newcommand{\sa}{\sigma_{\mathrm{a}}}

\newcommand{\rmin}{{\rm in}}
\newcommand{\rmex}{{\rm exc}}
\newcommand{\rmrem}{{\rm rem}}
\newcommand{\rmR}{{\rm R}}
\newcommand{\rmT}{{\rm T}}

\newcommand{\Iol}{I_{\mathrm{in}}(\lex)}

\newcommand{\Irem}{I^{\mathrm{R}}_{\rmrem}(\lex,\lem)}
\newcommand{\Item}{I^{\mathrm{T}}_{\rmrem}(\lex,\lem)}

\newcommand{\Y}{\textrm{YAG:Ce}^{+3}}
\newcommand{\lex}{\lambda_1}
\newcommand{\lem}{\lambda_2}

\newcommand{\FIG}[4]{
	\begin{figure}[t!]
		\centering
		\includegraphics{#2.jpg}
		\caption[#3]{\textbf{#3} #4}
		\label{#1}
	\end{figure}
}

\newcommand{\FIGefit}[4]{
	\begin{figure}[h!]
		\centering
		\includegraphics[width=\textwidth]{#2.jpg}
		\caption[#3]{\textbf{#3} #4}
		\label{#1}
	\end{figure}
}

\newcommand{\FIGesmal}[5]{
	\begin{figure}[h!]
		\centering
		\includegraphics[width=#3]{#2.pdf}
		\caption[#4]{\textbf{#4} #5}
		\label{#1}
	\end{figure}
}

\author{Maryna L. Meretska}
\affiliation[COPS]
{Complex Photonic Systems (COPS), MESA+ Institute for Nanotechnology, University of Twente, P. O. Box 217, 7500 AE Enschede, The Netherlands}
\author{Gilles Vissenberg}
\affiliation[COPS]
{Philips Lighting, High Tech Campus 7, 5656 AE Eindhoven, The Netherlands}
\author{Ad Lagendijk}
\affiliation[COPS]
{Complex Photonic Systems (COPS), MESA+ Institute for Nanotechnology, University of Twente, P. O. Box 217, 7500 AE Enschede, The Netherlands}
\author{Wilbert L. IJzerman}
\affiliation[COPS]
{Philips Lighting, High Tech Campus 7, 5656 AE Eindhoven, The Netherlands}
\alsoaffiliation[Eindhoven University]
{Department of Mathematics and Computer Science, Eindhoven University of Technology, 5600 MB Eindhoven, the Netherlands}
\email{wilbert.ijzerman@philips.com}
\author{Willem L. Vos}
\affiliation[COPS]
{Complex Photonic Systems (COPS), MESA+ Institute for Nanotechnology, University of Twente, P. O. Box 217, 7500 AE Enschede, The Netherlands}
\email{w.l.vos@utwente.nl}

\title[title]
  {Systematic design of the color point of a white LED}


\begin{document}


\begin{abstract}
  Lighting is a crucial technology that is used every day. 
  The introduction of the white light emitting diode (LED) that consists of a blue LED combined with a phosphor layer, greatly reduces the energy consumption for lighting. 
  Despite the fast-growing market  white LED's are still designed using slow, numerical, trial-and-error algorithms. Here we introduce a radically new design principle that is based on an analytical model instead of a numerical approach. 
  Our design model predicts the color point for any combination of design parameters. 
  In addition the model provides the reflection and transmission coefficients - as well as the energy density distribution inside the LED - of the scattered and re-emitted light intensities. 
  To validate our model we performed extensive experiments on an emblematic white LED and found excellent agreement. 
  Our model provides for a fast and efficient design, resulting in reduction of both design and production costs.
  
\end{abstract}


\noindent The main characteristics of white light sources are their color point~\cite{Malacara11} and efficiency. The color point is described using two independent chromaticity parameters that fill the color space~\cite{CIE30,Smith30}. 
The color point of any light source is defined by the spectrum it emits. 
Optical designers currently use slow numerical simulations, often based on a Monte-Carlo ray tracing techniques~\cite{Luo05,Tran08} to extract the color point given the design parameters of the white light source.

Conversely, to target a specific color point, optical designers have to use these slow simulations for each chosen set of design parameters when scanning the design parameter space. 

An important class of white light sources are the white LEDs that possess numerous advantages over conventional sources, such as incandescent lamps or discharge lamps. 
White LEDs are one of the most energy efficient sources~\cite{Almeida11,Danish16}, they are mechanically robust and thermally stable, they possess good temporal stability and they have a long lifetime~\cite{Editorial07,Pulli15}. 
To systematically design the color point of a white LED~\cite{Schubert06,Horiuchi10} algorithms are needed that are much faster than the ray tracing techniques.

A typical white LED consists of a blue semiconductor LED~\cite{Nakamura92,Amano89,Nakamura09} in combination with a phosphor layer~\cite{Schubert06}, that consists of a dielectric matrix with a density $\rho$ of phosphor micro particles (see Fig.~\ref{fig:sch}). 
Part of the blue light is transmitted through the phosphor layer, and part is absorbed and re-emitted in the red and green part of the spectrum to yield the desired white light to illuminate a targeted object or space. 
The relative amount of scattered and re-emitted light defines the color point of a white LED.
%
%
\FIGefit{fig:sch}{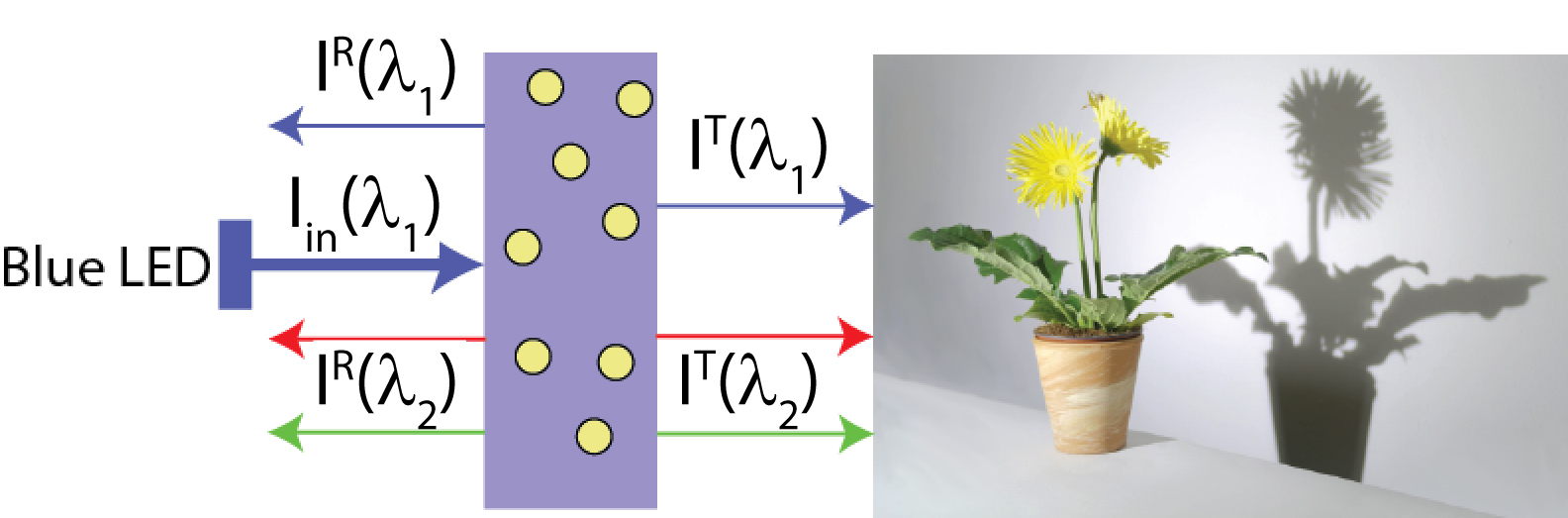}{Model of light propagation in a white LED slab.}{Blue excitation light with intensity $I_\rmin(\lex)$, originating from the blue LED, is shone on the phosphor slab with thickness L. 
The phosphor slab contains phosphor micro particles that are represented by yellow circles. 
$I^{\rmT}(\lex)$ is the scattered transmitted intensity, $I^{\rmR}(\lex)$ is the scattered reflected intensity. 
$I^{\rmT}(\lem)$ is the transmitted re-emitted intensity and $I^{\rmR}(\lem)$ is the reflected re-emitted intensity, both per bandwidth of the detector at $\lem$. 
The mixture of transmitted red, green and blue light illuminates the object.
}
%
To adjust the color point several design parameters, such as the phosphor layer thickness $L$, the phosphor particle density $\rho$, the phosphor type, the type of blue LED, the particle density of the additional scattering elements are available.
In this paper, we introduce an extremely fast, analytic computational tool - based on the so-called P3 approximation (see Methods) - to predict the color point of a white LED starting from the chosen design parameters, or inversely, to infer the design parameters of a white LED beginning from a targeted color point. 
In our case the inverse problem does not require an iteration procedure for each new design cycle. 
Given the speed of our tool, we can generate -~once and for all~- a look-up table for the whole parameter space available to engineers. 

The core of our method is to calculate the spatial light energy distribution of the scattered and of the re-emitted light inside a white LED.
From this spatial profile the reflection and transmission coefficients can be obtained. 
Studying the sensitivity of the energy distribution to changing design parameters gives much insight and can be used to design more efficient and more robust LEDs. 
For instance, it was recently demonstrated that the functioning of phosphor layers could degrade substantially due to thermal effects~\cite{Tan18}. 
In addition heat generated from the phosphor particles (Stokes losses) can cause  damage to the polymer matrix~\cite{Singh16}.
Degradation due to heating can be explained by inspecting the internal energy distribution, and can even be prevented by choosing design parameters that avoid hot spots. Nonlinear effects, caused by peaks in the energy profile, lead to quenching and lifetime degradation of a white LED. 
For high-power laser-driven phosphor-based white LEDs that are used for outdoor lighting or car headlamps~\cite{Song16}, engineering of the energy distribution is crucial to minimize the effect of quenching~\cite{Kim17}. 
In more exotic cases, like quantum dot based or dye-based white LEDs~\cite{DiMartino14}, hot spots cause bleaching of the quantum dots or the dye. 

Our analytical model generates all light fluxes - scattered and re-emitted - emanating from a white LED at all wavelengths (see Fig.~\ref{fig:sch}). 
In addition the corresponding energy densities inside the LED are a natural result of our approach.
%
%
%
The transmission $T_{\rmex}(\lex)$ and reflection $R_{\rmex}(\lex)$ coefficients of the excitation light that is partly scattered are analytically calculated using the P3 approximation to the radiative transfer equation (see Methods).
To solve the P3 approximation for the scattered light the following properties of the phosphors are needed: the scattering cross section $\sss (\lex)$, the absorption cross section $\sa (\lex)$, the anisotropy factor $\mu (\lex)$, the thickness of the phosphor layer $d$ and the particle density $\rho$ of the phosphor particles.
The concentration and the slab thickness are chosen by the optical designer and the other three parameters can be obtained for the whole visible spectral range from experiments as described elsewhere~\cite{Vos13,Gaonkar14,Meretska17} or using Mie theory~\cite{Bohren83}. 
We obtain the spatial energy distribution $U_{\rm exc}(\lex,z)$ of the excitation light inside the slab~\cite{Ishimaru78}, the transmission and reflection coefficients with P3 approximation.
%
%
 To obtain the re-emitted intensities the average energy density of the excitation light $U_{\rmex}(\lex,z)$ is used as a source function for the transport of the re-emitted light. 
 At position $z$ the excitation intensity is proportional to $U_{\rmex}(\lex,z)$ and the power of the re-emitted light at wavelength $\lem$ is proportional to $q(\lex) U_{\rmex}(\lex,z) F(\lem)$,  where  $F(\lem)$ is the normalized re-emission (fluorescence)  spectrum, and where $q(\lex)$ is the quantum efficiency. 
Both the normalized re-emission spectrum and the quantum efficiency can be extracted from experiments and are known for many phosphors~\cite{Smet11}.

The scattering of the re-emitted light by a particle is characterized by the scattering cross section $\sss(\lem)$ the anisotropy factor $\mu(\lem)$, and the absorption cross section $\sa(\lem)$. 
Application of the P3 approximation to the transport of the re-emitted intensity generates both the energy density of the re-emitted light, $U_{\rem}(\lem,z)$, and also the transmission and the reflection of the re-emitted light. 
We introduce the differential re-emitted transmission intensity $\Item d \lem$ and reflection intensity $\Irem d \lem$ and use them to define the total re-emitted intensities by integrating them over the re-emission spectrum:  
$ \int \Item d \lem$ and $\int \Irem d \lem$. 
When normalizing the total re-emitted intensities with the incident intensity we obtain dimensionless coefficients that can be easily compared to experiments:
$T_{\rem}(\lex)\equiv \int \Item d \lem/\Iol$, 
and
$R_{\rem}(\lex)\equiv \int \Irem d \lem/\Iol$.

\section{Results and Discussion}
%
To demonstrate the validity of our tool we experimentally study light propagation through a representative white LED and interpret the results using our model. 
The LED consist of a polymer diffuser plate containing $\Y$ phosphor micro particles. 
The refractive index of the diffuser plates is $n=1.4$, and the plate thickness $L=1.98\pm 0.02$~mm. 
The absorption and the emission spectra of the phosphor are discussed elsewhere~\cite{Meretska16}. 
The phosphor particle density $\rho$ ranges from 1~wt$\%$ to 8~wt$\%$. 
A narrow-band light source, tunable from 420 nm to 800 nm, with intensity $\Iol$ illuminates the samples at a wavelength $\lex$~\cite{Meretska16}.
The incident light is scattered, absorbed, and re-emitted. 
The transmitted and reflected intensities are separately collected with an integrating sphere and detected with a fiber spectrometer. 
%
%
 \FIGesmal{fig:set}{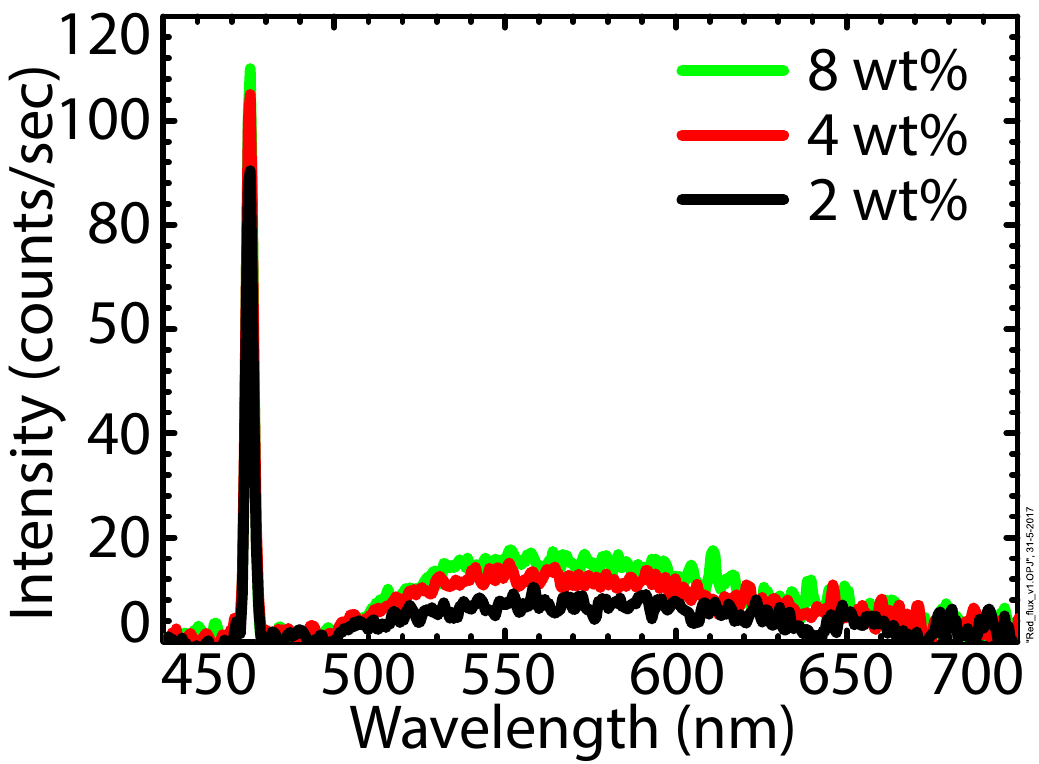}{4 in}{Typical measured signal of the model white LED.} { Observed emission from a white LED in reflection when excited with $\lex=475$ nm narrow-band light, for three different phosphor concentrations.  
 The sharp peak at 475~nm is the scattered incident light in reflection and the broad peak between 490 nm and 700 nm is light re-emitted by the phosphor. }
%
%
The typical signal of our model white LED is shown in Fig.~\ref{fig:set}. The sharp peak at $\lex = 475$~nm originates from the light source and contributes to the scattered light intensities. 
The broad peak between 490 nm and 700 nm consists of the re-emitted  light. 

From the experiment we infer the following quantities that are compared with the predictions of our model: 
the scattered transmission coefficient $T_{\rmex}(\lex)$, the scattered reflection coefficient $R_{\rmex}(\lex)$, the total transmission coefficient of re-emitted light $T_{\rem}(\lex)$, and the total reflection coefficient of re-emitted light $R_{\rem}(\lex)$. 
The total re-emitted transmission coefficient $T_{\rem}(\lex)$ and the total re-emitted reflection coefficient $T_{\rem}(\lex)$ are obtained by integrating the differential transmission and reflection coefficients over $\lem$ from 490~nm to 700~nm by binning the wavelength in 841 points.
%
%
\FIGefit{fig:TR}{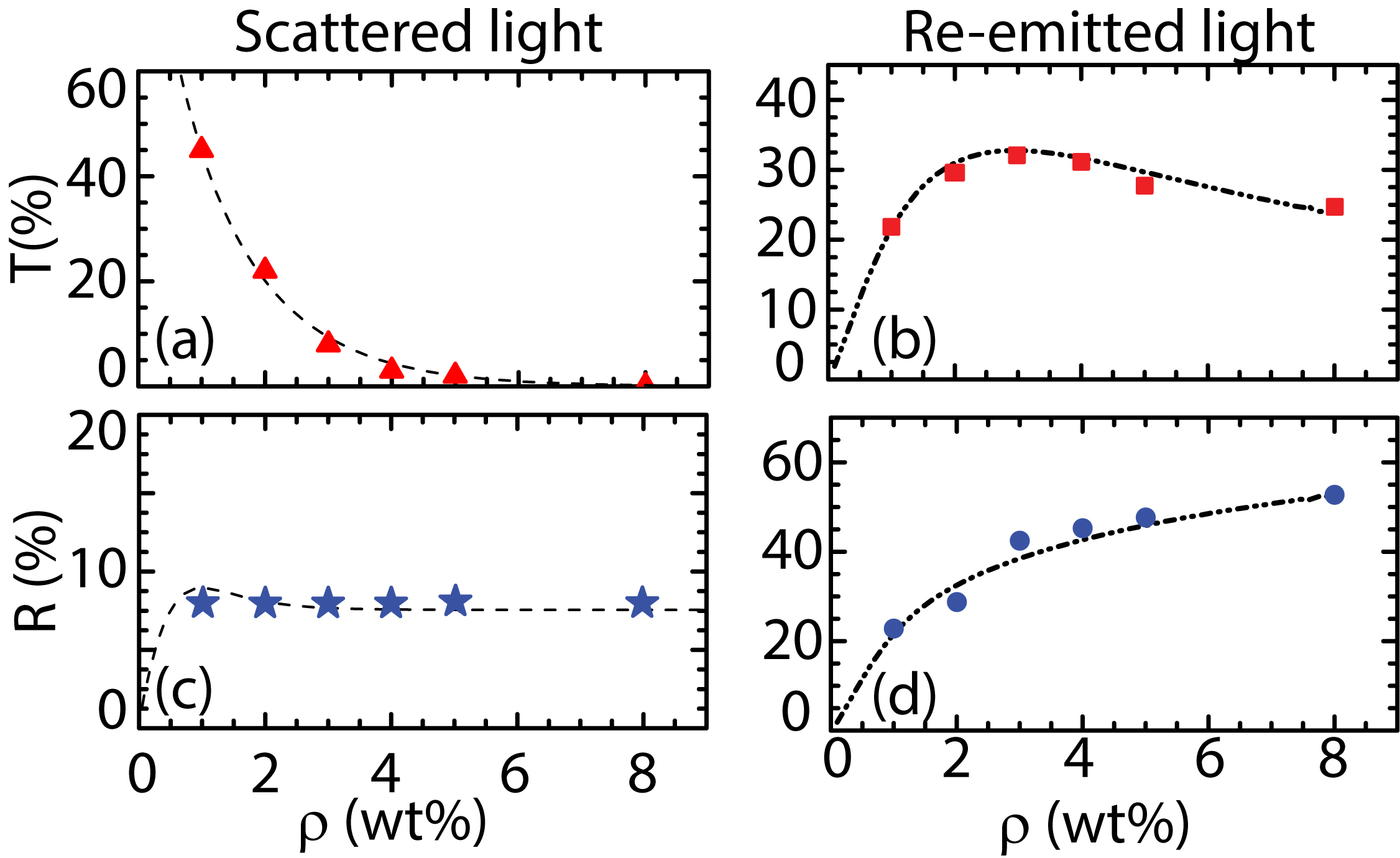}{Transmission and reflection of a model white LED as a function of phosphor particle density at $\lex=475$~nm}
{\textbf{(a)}~Dashed line represents the calculated total transmission coefficient of the scattered light. Triangles represent the measured coefficient, 
\textbf{(b)}~dash-dot-dot line represents the calculated total transmission coefficient of the re-emitted light. Squares represent the measured coefficients, 
\textbf{(c)}~dashed line represents the calculated reflection coefficient of the scattered light. Stars represent the measured coefficient,
\textbf{(d)}~dashed-dot line represents the calculated reflection coefficient of the re-emitted light. Circles represent the measured 
coefficient. The error bars of the experiment is within the symbol size. 
}
%
%
%
The experimental results for the scattered blue light are shown in Fig.~\ref{fig:TR}. The transmission $T_{\rmex}(\lex)$ of the scattered light reveals a sharp decrease with the phosphor particle density $\rho$ as a result of scattering and strong absorption of the incident light. The reflection $R_{\rmex}(\lex)$ hardly varies with the phosphor particle density due to absorption; if the phosphor did not absorb light, we would have observed growth of the reflection with the increasing phosphor particle density, but the absorption counters this growth.
The measured total transmission coefficient $T_{\rem}(\lex)$ and total reflection coefficient $R_{\rem}(\lex)$ of the re-emitted light are also shown in Fig.~\ref{fig:TR} (a)-(b). At low particle density, the re-emitted light intensity $R_{\rem}(\lex)$ increases with the phosphor particle density, because more phosphor absorbs more light. 
A saturation occurs at $\rho=3.3$~wt$\%$, because a further increase of the particle density increases the probability of incident photons being absorbed very close to the entrance surface of the scattering material. 
When most photons are absorbed near the entrance surface, most re-emitted photons leave through the entrance surface, corresponding to a decrease of the transmitted re-emitted flux, and an increase of the reflected re-emitted flux.
The color points for the measured data are plotted in the color space in Fig.~\ref{fig:cp-data}. 
%
%
%
\FIG{fig:cp-data}{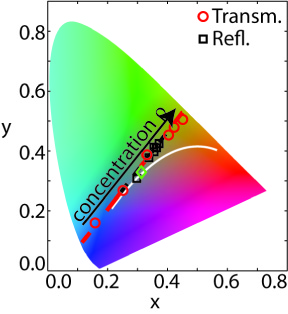}{Color point of a white LED.}{
Circles (transmission) and squares (reflection) are our experimental data points.
Red and black dashed lines represent predicted color point dependence (see curves in Fig.~\ref{fig:TR}) on the phosphor {particle density} $\rho$ (1 wt$\%$ to 8 wt$\%$) for transmitted and reflected light, respectively. 
 The green diamond indicates the most widely used standardized white light spectrum - D65 spectrum. 
The white curve represents the Planckian locus of black-body sources. 
{The experimental data points are listed in the Supporting Information.}}
%
%
The experimentally determined quantum yield can be obtained 
from $q(\lex) = \{T_{\rem}(\lex)+R_{\rem}(\lex)\}/\{1-T_{\rmex}(\lex) -R_{\rmex}(\lex)\}$ and is presented in Fig.~\ref{fig:q}.

%
\FIG{fig:q}{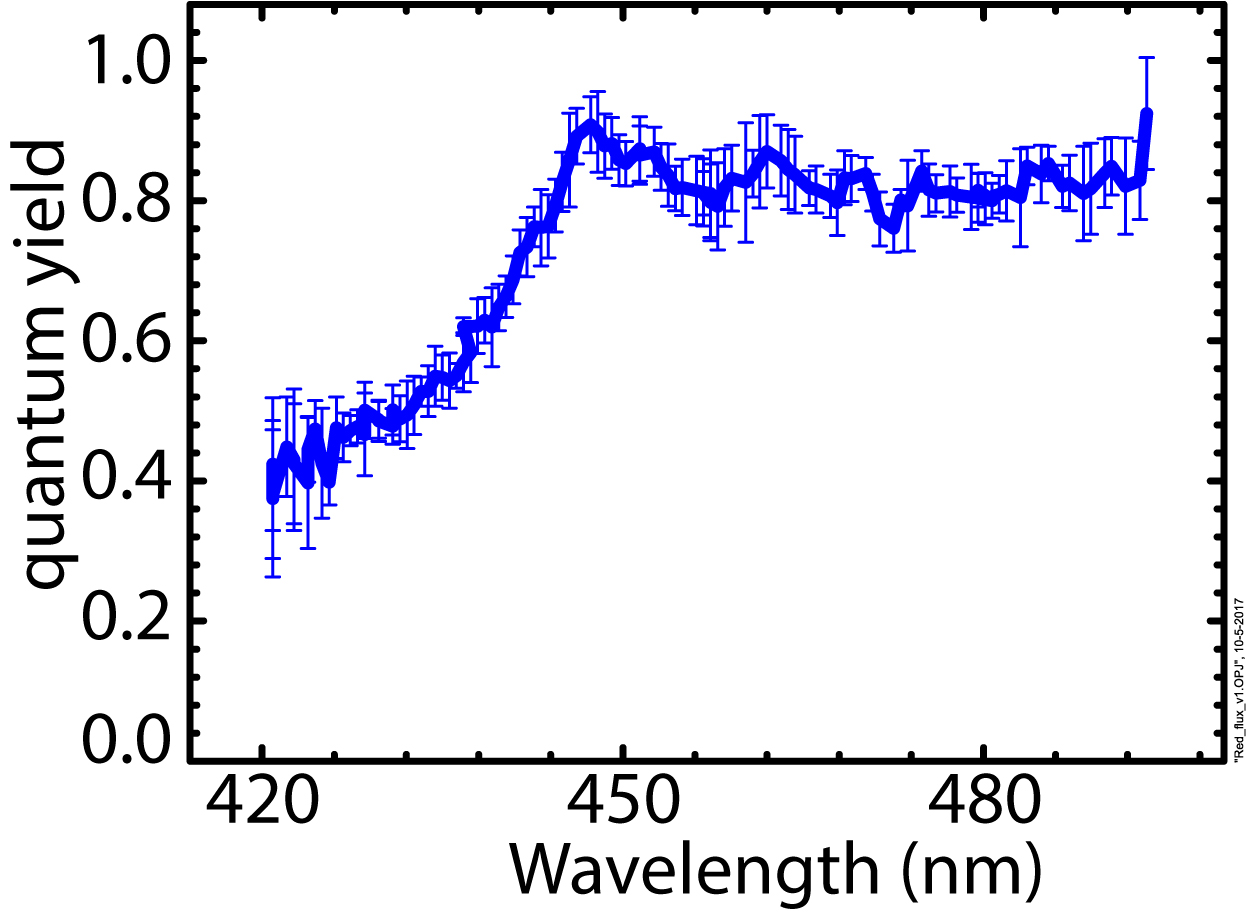}{Quantum yield of $\Y$.}{Quantum yield as a function of the excitation wavelength $\lex$ as extracted from our measurements.}
%
%
%
\section{Model validation}
%
The results of our analytical model for the transmissions $T_{\rmex}(\lex)$ and $T_{\rem}(\lex)$ and the reflections $R_{\rmex}(\lex)$ and $R_{\rem}(\lex)$ as function of particle density $\rho$ are also shown in Fig.~\ref{fig:TR} (a)-(b) together with our experimental data. 
We find an excellent agreement between experiment and our model. 
To explain the behaviour of the $T_{\rem}(\lex)$ and $R_{\rem}(\lex)$ as a function of phosphor particle density we plot the absorbed power $4\pi \rho \sa U_{\rmex}(z)$ as a function of depth inside the scattering medium in Fig.~\ref{fig:UZ}(c) where the $z$ coordinate corresponds to the direction in Fig.~\ref{fig:UZ}.
When the particle density of the phosphor particles is low the absorbed power is distributed almost uniformly across the sample. Re-emitted light generated as a result of absorption in a representative layer {$dz_\mathrm{B}=L/3$} contributes mostly to the re-emitted reflection $R_{\rem}(\lex)$, because light that is re-emitted in the backward direction in this layer will experience less scattering compared to the forward re-emitted light. 
The re-emitted light in a second representative layer {$dz_\mathrm{F}=2L/3$} will mostly contribute to the transmission $T_{\rem}(\lex)$ for the same reason. When the phosphor particle density is high the intensity is mostly absorbed near the entrance surface of the slab, hence layer $dz_\mathrm{B}$ will contribute a higher intensity in the backward direction compared to the low particle density case, and the layer $dz_\mathrm{F}$ will contribute less intensity in transmission $T_{\rem}(\lex)$ compared to the low particle density case. 
This interplay results in the peak in transmission $T_{\rem}(\lex)$ at $\rho=3.3$~wt$\%$ and a steady increase of reflection $R_{\rem}(\lex)$ with particle density $\rho$.
%
%
\FIG{fig:UZ}{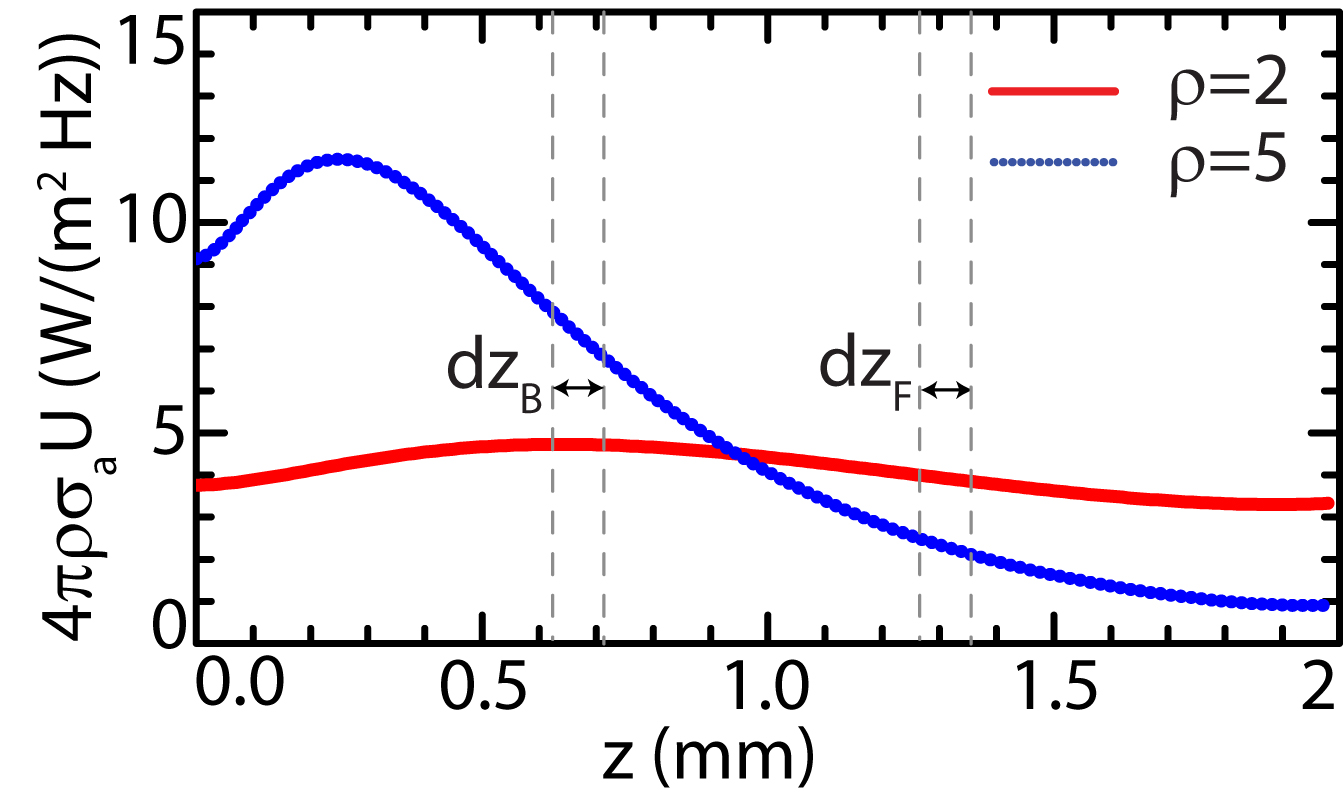}{Energy density profile inside model white LED}
{
Lines represent the calculated average intensity $U(z)$ of the blue light as a function of depth inside the sample for two different phosphor particle particle density $\rho$. 
Two representative ranges highlighted with grey dashed lines are showed at L/3 and 2L/3 depth in the sample and discussed in the text.}
%
%

The color points for the measured data and our model are plotted in the color space in Fig.~\ref{fig:cp-data} and show excellent agreement.
The theoretical curves indicate all possible color points that can be achieved with the given $\Y$ phosphor, when the particle density $\rho$ is changed. 
The observed dependence of the color point on particle density is remarkably linear given the observation in Fig.~\ref{fig:TR} that the scattered and re-emitted light intensities depend non-linearly on the particle density. 
The color point linearly shifts to the yellow part of the color space with increasing particle density for both the transmitted and reflected light as shown in Fig.~\ref{fig:cp-data}. 
The reason is that a higher particle density enhances the absorption of blue light increasing the produced re-emitted light.  
The re-emitted flux has qualitatively an identical dependence on both the particle density and the thickness of the polymer layer. 
When the refractive index of the polymer changes from 1.4 to 1.5 typical for industrially used materials, the re-emitted flux increases by no more than $4\%$. 
The widely used design of a white LED typically contains a mirror that is used to reflect backscattered light flux~\cite{Schubert06}. 
Our model can be easily extended to include this mirror. 
We considered a silver mirror with a reflectivity of 99\% and assuming the quantum efficiency of the phosphor to be $q=1$. 
When a mirror is introduced in the design the amount of the re-emitted flux grows by up to 40\% compared to the design without a mirror. 
The scattered flux, however, changes insignificantly, the change is less than 2\%.

%
%
\section{Conclusions}
%
%
We have developed an analytic and very fast model that can revolutionize white LED design. It provides a simple, efficient design tool allowing to access a large design parameter space using an analytic algorithm without adjustable parameters. 
Our design principle is based on calculating the spatial light energy density inside the LED. This property is crucial in designing a white LED, as it allows not only to predict the color of a white LED, but also supplies information about prevention the detrimental effects of heat generation inside the phosphor layer on the layer itself and on the polymer host. 
Our design method can be applied not only to design traditional phosphor-based white LEDs, but also white LEDs that are a topic of quickly developing research fields like laser-driven white LEDs or quantum dots white LEDs.
While our current solution provides a single-layer description of the light propagation and re-emission, the extension to multiple layers, but without the serious complication of re-emission, has already been reported~\cite{Liemert17}  and the extension to more complex geometries 
typically employed in white LED design seems realizable. Our approach will increase the design efficiency by avoiding recurring design efforts and decrease the cost of ownership of white LEDs units for worldwide users, and is already being put to use by engineers in industry. 
%
\section{Methods}
%

\subsection{P3 approximation}
%
The key property to calculate in radiative transport is the specific intensity $I( \bm r , \bm {\hat{s}})$, which for direction $\bm {\hat{s}}$ at position $\bm r$ represents the average power flux density per unit frequency and per unit solid angle. 
To separate the coupled dependency on the variables $\bm r$ and $ \bm {\hat{s}}$, one expands into a product of functions that depend on either $\bm r$ or $ \bm {\hat{s}}$
\begin{equation}
I( \bm r , \bm { \hat{s}}) =\sum\limits_{l=0}^{\cal L} \sum\limits_{m=-l}^{l} \psi_{lm}(\bm r) Y_{lm}(\bm {\hat{s}})
\mathrm{,}
\label{eq:pn}
\end{equation}
where $Y_{lm}$ are the spherical harmonics and the functions $\psi_{lm}$ are determined by the boundary conditions. 
In principle the expansion is exact when we take $\cal L$ to infinity. When limiting $\cal L$ to finite N, the expansion is called the PN approximation. 
It appears that only odd N gives sensible results.
In media with strong absorption the P1 approximation  - also called the diffusion approximation - is known to fail~\cite{Meretska17}. 
For the diffusion approximation to be valid, the spatial gradients of the specific intensity have to be small. 
With strong absorption, this gradient becomes too large as absorption induces an exponential decay with a decay length less than the scattering mean free path. 
In these cases, one has to resort to the P3 approximation~\cite{lie16,Star89,Dickey98,Klose06} to the radiative transfer equation~\cite{Chandrasekhar60}. 
Only in exceptional cases, characterized by high absorption combined with high scattering anisotropy, higher approximations - like the complex P5 approximation - have to be invoked.
The derivation of the P3 approximation for several geometries has been published before~\cite{lie16,Star89,Dickey98,Klose06,Chandrasekhar60}. The formula's for the P3 approximation can be obtained by simply using symbolic manipulation by Mathematica. 
Since the P3 equations for the slab geometry have not been explicitly published as far as we know, we fully provide them in the Supplementary Information.  
%
\subsection{Comparison with alternative models}
%
The alternative computational methods that can compete with our model regarding accuracy but not regarding speed are Monte-Carlo simulations and the adding-doubling method~\cite{Hul80I,Prahl93}. 
The adding-doubling method consists of approximating the transport properties of a very thin, hypothetical, layer very accurately and then further doubling the thin layer again and again until the required layer thickness is reached. 
Any doubling cycle requires numerical matrix inversion and angular integrals, approximated by discretizing the angular coordinates and using a numerical quadrature method like Gaussian-Legendre. 
The adding-doubling method has been very successful in describing the optical properties of biological tissue~\cite{Prahl93}. 
To apply the adding-doubling method to light transport in LEDs it has to be generalized to include the substantial complication of re-emission~\cite{ley12}. 
Very recently this more complex adding-doubling method was applied to LED design~\cite{ryc15}. 
In this generalized form the numerical calculation of transmission and reflection has to be done for each excitation wavelength $\lex$ and for each re-emission wavelength $\lem$. 
According to Leyre et al.~\cite{ley12} doing a ray tracing calculation on a single $\{\lex,\lem \}$ pair would take several hours on a standard PC, whereas the adding-doubling method takes only about 35 seconds. 
However, for the design of a LED the light transport for the whole re-emission spectrum has to be calculated, implying  the discretization of $\lem$ in at least 100 points. 
This full spectral scan of the re-emitted light with the adding-doubling method would take of the order of an hour.
With our analytic method on the other hand, where we bin $\lem$ in no less than 841 points, the {\it whole} scattering and re-emission calculation takes - with Mathematica - about 0.75 seconds. 
As a matter of fact the integration over the re-emission spectrum only implies a small additional overhead, and even a full excitation scan combined with each time a full re-emission scan would take only a few minutes. 
To obtain the energy distribution inside a LED using the adding-doubling method is {\it in principle} possible, but cumbersome and has - to the best of our knowledge - not yet been reported.

\section{Author Contributions}
M.L.M. performed the experiments with assistance from W.L.V., G. V., W.L.IJ.;
M.L.M., and W.L.V. analyzed the data;
M.L.M. and A.L. developed the theory; 
W.L.V., A.L., and W.L.IJ. supervised the project. 
All authors wrote the manuscript.

\begin{acknowledgement}
It is a great pleasure to thank Jan Jansen from Philips Lighting for sample fabrication, Cornelis Harteveld for technical support, Diana Grishina, Oluwafemi Ojambati, Ravitej Uppu, and Shakeeb Bin Hasan for useful discussions, and Nono Groenen for helping with Fig.~\ref{fig:sch} and the TOC figure. 
This work was supported by the Dutch Technology Foundation STW (contract no. 11985), and by the NWO-FOM program "Stirring of Light!", by the Dutch Funding Agency NWO, and by MESA+ Applied Nanophotonics (ANP).
\end{acknowledgement}

\begin{suppinfo}

\end{suppinfo}


\bibliographystyle{achemso}
\bibliography{References_color}

\end{document}